\documentclass[12pt,preprint]{aastex}

\def\gs{\mathrel{\raise0.35ex\hbox{$\scriptstyle >$}\kern-0.6em
\lower0.40ex\hbox{{$\scriptstyle \sim$}}}}
\def\ls{\mathrel{\raise0.35ex\hbox{$\scriptstyle <$}\kern-0.6em
\lower0.40ex\hbox{{$\scriptstyle \sim$}}}}

\shorttitle{Warm molecular gas in spiral disks}
\shortauthors{Papadopoulos, Thi, \& Viti}

\begin{document}

\title{Molecular gas in spiral galaxies: a new warm phase at large
galactocentric distances?}

\author{P.\ P.\ Papadopoulos}
\affil{ESA Astrophysics Division, Research and Scientific Support Department,
             ESTEC, Postbus 299, 2200 AG Noordwijk,
             The Netherlands}

\author{W.-F.\ Thi}
\affil{Department of Physics \& Astronomy, University College
                London,  London WC1E 6BT}

\and

\author{S. Viti}
\affil{Department of Physics \& Astronomy, University College
                London,  London WC1E 6BT}

\begin{abstract}

There is  now strong  evidence suggesting that  the $  ^{12}$CO J=1--0
transition,   widely   used   to   trace  H$_2$   gas,   significantly
underestimates its  mass in metal-poor  regions. In spiral  disks such
regions are found in large galactocentric distances where we show that
any unaccounted  H$_2$ gas phase is  likely to be  diffuse ($\rm n\sim
5-20$~cm$ ^{-3}$) and warmer ($\rm T_{\rm kin}\sim 50-100$ K) than the
cool  ($\rm T_{\rm  kin}\sim~15-20$~K) CO-luminous  one.   Moreover we
find  that  a  high value  of  the  H$_2$  formation rate  on  grains,
suggested  by  recent  observational  work,  can  compensate  for  the
reduction of  the available  grain surface in  the metal-poor  part of
typical galactic  disks and thus enhance this  CO-poor H$_2$ component
which may  be contributing significantly  to the mass and  pressure of
spiral disks beyond their optical radius.

\end{abstract}


\section{Introduction}

The use of $ ^{12}$CO,  the most abundant molecule after H$_2$ itself,
to trace  H$_2$ gas is a  widely used practice since  the discovery of
bright  $ ^{12}$CO  J=1--0  emission in  Orion  (Wilson, Jefferts,  \&
Penzias  1970).   The  most   easily  excited  H$_2$  transition,  the
quadrupole  S(0): $\rm J_{u}-J_{l}=2-0$,  corresponds to  $\Delta \rm
E_{20}/k\sim 510$  K, much too  high to be significantly  populated in
the  cool  ISM ($\rm  T_{\rm  kin}\sim 10-50$  K)  revealed  by the  $
^{12}$CO  J=1--0   transition  ($\Delta  \rm   E_{10}/k\sim  5.5$  K).
Furthermore the  large optical depths  of the latter  ($\tau _{10}\sim
5-10$)  reduce its  critical density  of $\rm  n^{(10)}  _{\rm cr}\sim
380$~cm$ ^{-3}$  (for $\rm T_{\rm kin}=50$~K; Kamp  \& Zadelhoff 2001)
to  $\sim  \rm  n^{(10)}   _{\rm  cr}(1-e^{-\tau  _{10}})  \tau  ^{-1}
_{10}\sim  40-80\ cm^{-3}$,  similar to  the lowest  average densities
observed  in Giant  Molecular  Clouds (GMCs).   Hence  the $  ^{12}$CO
J=1--0 transition  is expected to be  excited even in  the coldest and
most diffuse regions of the molecular~ISM.

However  its large  optical depths  (e.g.  Martin,  Sanders,  \& Hills
1984;  Sage  \&   Isbell  1991;  Falgarone  1998),  do   not  allow  a
straightforward  use  of  its   luminosity  together  with  a  $\rm  [
^{12}CO/H_2]$ abundance  ratio for  H$_2$ mass estimates.   The widely
used $\rm X_{\rm CO}=M(H_2)/L_{\rm CO}$ factor (where $\rm L_{\rm CO}$
is the  velocity/area-integrated brightness temperature  of $ ^{12}$CO
J=1--0)  is based  on  the  notion of  an  effectively optically  thin
medium, where the $ ^{12}$CO  line emission arises from an ensemble of
small,  radiatively  de-coupled  cells  not overlapping  in  space  or
velocity.   Then the  high  optical depths  arise  locally within  the
individual cells and the observed wide line profiles result from their
macroscopic  motions  rather   than  their  intrinsic,  much  narrower
linewidths  (e.g.  Martin  et al.   1984; Young  \& Scoville  1991 and
references   therein).   This   picture,  along   with   the  observed
virialization of molecular clouds (see  e.g.  Larson 1981) over a wide
range of scales ($\sim 0.1-100$ pc) allows a statistical derivation of
$\rm X_{\rm CO}$ as an ensemble-average that is relatively insensitive
to local molecular cloud conditions (e.g.  Dickman, Snell, \& Schloerb
1986; Young \& Scoville 1991; Bryant \& Scoville 1996) and, for scales
$\ga 10$ pc,  is expected to yield the correct H$_2$  mass to within a
factor of two.

The existence of such cells  is supported by high angular and velocity
resolution studies  of galactic clouds (e.g.  Falgarone  et al.  1998;
Tauber 1996) while the statistical  robustness of $\rm X_{\rm CO}$ for
a  variety of conditions  has been  verified using  both observational
(e.g.  Young \& Scoville 1991 and references therein), and theoretical
methods (e.g. Kutner  \& Leung 1985; Dickman, Snell  \& Schloerb 1986;
Maloney \& Black 1988;  Wolfire, Hollenbach, \& Tielens 1993; Sakamoto
1996; Bryant \& Scoville 1996).   However these studies also suggest a
likely failure  of this method in two  particular environments, namely
a) galactic and/or starburst nuclei where non-virial motions cause the
luminous  $ ^{12}$CO  emission to  overestimate the  H$_2$  mass (e.g.
Dahmen et  al.  1998; Downes  \& Solomon 1998), b)  metal-poor regions
where  it seriously underestimates  the H$_2$  mass (e.g.   Maloney \&
Black  1988; Arimoto,  Sofue, \&  Tsujimoto 1996;  Israel  1988, 1997,
1999).  The  last case is the  most serious one since  then $ ^{12}$CO
emission is usually faint and  thus precludes the observations of more
optically thin isotopes  like $ ^{13}$CO, C$ ^{18}$O  that could yield
the H$_2$ mass without using the $\rm X_{\rm CO}$ factor.

In  this   paper  we  show   that  the  decreasing   metallicity  with
galactocentric radius in disks and  a larger H$_2$ formation rate open
up the possibility of large  amounts of H$_2$ gas being there hitherto
undetected  by the  conventional  method.  This  phase  will be  warm,
diffuse  and devoid  of CO,  and may  contribute significantly  to the
pressure and total mass of the disk.

\section{Spiral disks: H$_{\bf 2}$ formation in the metal-poor parts}

Molecular   hydrogen  in   interstellar  conditions   forms   from  HI
association onto dust  grains (e.g.  Gould \& Salpeter  1963) and deep
sub-mm  imaging  shows the  grain  distribution  extending beyond  the
optical and  well into the HI  disk (Nelson, Zaritsky,  \& Cutri 1998;
Xylouris et al.  1999), and possibly out to $\sim 2\rm R_{25}$ (Thomas
et al.   2001, 2002;  Guillandre et al.   2001).  {\it Thus  the basic
``ingredients''  for H$_2$  formation can  be found  to  the outermost
parts of the disk as defined by the HI distribution itself.} Using the
results of  Federman, Glassgold, \&  Kwan (1979), with an  exponent of
m=1.5 for the H$_2$ self-shielding  function, we find that an HI cloud
of  radius  $\rm  R_{\rm  cl}$  (pc), average  volume  density  n  and
temperature  T, illuminated by  a FUV  field strength  $\rm G_{\circ}$
($\rm G_{\circ}$(solar)=1) starts turning molecular when $s>1$, where

\begin{equation}
s=\frac{\rm n}{\rm n_{\rm min}}=\rm R_{\rm cl} ^{2/5}\left(\frac{\rm n}{65\rm 
cm^{-3}}\right) \left(\frac{\rm T}{100\rm K}\right)^{3/10}\left(
\frac{\mu \rm z}{\rm G_{\circ }}\right)^{3/5}.
\end{equation}

\noindent
The density $\rm n_{\rm min}$ is the minimum required for the onset of
an HI-to-H$_2$  transition and $\rm z= Z/Z_{\odot}$  is the normalized
metallicity.  We kept the explicit temperature dependence of the H$_2$
formation  rate  $\rm R_{f}=S_{f}  T^{1/2}$,  assumed  that it  scales
linearly with  total grain surface and thus  metallicity, and included
the factor $\mu = \rm  S_{f}/S^{(o)} _{f}$, to parametrize for a value
different  than the  currently  adopted $\rm  S^{(o)}  _{f} =  3\times
10^{-18}$ cm$ ^{3}$ s$ ^{-1}$ K$ ^{-1/2}$ (Jura 1975a).

Over an HI disk it is $\rm T(HI)\sim 80-150$ K, with a slight increase
with  galactocentric  distance $\rm  R_{gal}$  (e.g.   Braun 1997),  a
temperature range  typical of the warm neutral  halos around molecular
clouds (e.g.  Andersson \& Wannier 1993) where the HI-H$_2$ transition
zone lies.  Thus  no significant variation of $s$  is expected because
of a  strongly varying  $\rm T(R_{gal})$.  The  influence of  the $\rm
z/G_{\circ  }$   ratio  is  less   straightforward  since  significant
metallicity gradients exist in spiral disks (e.g.  Henry 1998; Garnett
1998  and references therein)  along with  the well  known exponential
optical  (and hence  FUV)  light profiles.   Their  effects have  been
thoroughly  modeled by Elmegreen  (1989, 1993)  and applied  to spiral
galaxies  by Honma,  Sofue, \&  Arimoto (1995).   The latter  find the
metallicity gradient to  be the main cause of  the so called molecular
front, the  steep H$_2$-to-HI transition at a  particular $\rm R_{tr}$
deduced for  most spirals from  CO imaging.  Evidently a  larger H$_2$
formation rate  will alter this  picture by ``pushing'' this  front to
encompass a  larger portion  of a typical  disk. Rates that  are $\sim
4-8$  times  higher  than  the  standard  value  are  compatible  with
observationally deduced  upper limits  (Jura 1974), since  they always
contain  an  $\rm  n-S_f$  ``degeneracy''  that  can  bias  $\rm  S_f$
estimates towards  lower values (e.g.   Jura 1975a,b; van  Dishoeck \&
Black  1986).  Recent  results from  {\it ISO}  observations  of H$_2$
rotational lines from photodissociation  regions (Habart et al.  2000;
Li et  al.  2002) strongly favor  $\mu = \rm  S_{f}/S^{(o)} _{f}\ga 5$
(see also Sternberg  1988 for early indications of  such high values),
which will raise the value of $s$ at all $\rm R_{gal}$.

This rescaling has  no effect for those regions of  the disk where the
 old values  are already $s>1$ and  which CO imaging shows  them to be
 indeed H$_2$-rich, but  now a larger portion of  the disk is expected
 to have  $ s >  1$ and thus  conditions favorable for  an HI-to-H$_2$
 transition.  A rough estimate of this new radius can be obtained from
 the  results of Honma  et al.   (1995) that  show this  transition to
 occur very  sharply at a  particular $\rm R_{tr}$.  The  expected FUV
 volume emissivity  $\rm j_{\circ }$, its  dust absorption coefficient
 $\rm   k_{d}$,  and   the  metallicity   follow  a   similar  spatial
 distribution  (Honma  et  al.   1995)  hence,  neglecting  the  H$_2$
 self-shielding  contributions  to  the total  absorption  coefficient
 corresponds to $\rm  z/G_{\circ}=z/(j_{\circ}/k_{d}) \propto z$.  For
 an exponential disk profile and $\rm s(R_{tr})\sim 1$ it is

\begin{equation}
\rm R_{tr}(\mu )=R_{tr}(1)+(ln\mu) R_{e},
\end{equation}

\noindent
where $\rm R_{e}$ is  the scale-length of the exponential distribution
of optical light  and metallicity.  Thus the molecular  front will now
reside at least $\rm (ln5) \rm R_{e}\sim 1.6 R_{e}$ further, past that
inferred  by  CO imaging.   In  typical disks  the  later  is at  $\rm
R_{tr}(1)=(0.5-1) R_{e}$  (e.g.  Young \& Scoville 1991;  Regan et al.
2001), thus $\rm R_{tr}(\mu  )\sim (2-2.5) R_{e}$ places the molecular
front well  inside a typical HI  disk.  A flattening  of the abundance
gradient, and thus  of dust grain surface per H  nuclei, at large $\rm
R_{gal}$ (Henry  1998) and  H$_2$ self-shielding will  only strengthen
the  above arguments by  yielding still  larger values  of $s$  in the
outer parts of the disk.

Therefore it  now seems that  the molecular front in  spirals, deduced
from CO imaging, is in reality  a (low) metallicity-induced $ ^{12}$CO
detection  threshold,  and {\it  the  necessary  conditions for  H$_2$
formation exist  over a  much larger part  of a typical  spiral disk.}
This is not  surprising since it is known  that $\rm X_{\rm CO}\propto
z^{-1}-z^{-2}$ (e.g.  Arnault et  al.  1988; Israel 1997), which alone
implies  a $\rm R_{tr}$  larger than  that found  by using  $ ^{12}$CO
imaging and  a constant value of $\rm X_{\rm CO}$.

\subsection{The role of pressure in the HI/H$_2$ transition}

In obtaining the previous result  we assumed no significant changes of
 $\rm (n,  R_{cl})$ of the various  clouds with $\rm  R_{gal }$, which
 leaves only the  $\rm z/G_{\circ }$ variable to  define $\rm R_{tr}$.
 In a  forthcoming paper we will  present a more  detailed analysis of
 the   HI-H$_2$  transition  without   such  simplifications,   yet  a
 re-formulation of $s$ in terms  of cloud mass M and boundary pressure
 $\rm  P_e$  suggests  that  the  aforementioned  assumptions  do  not
 subtract from  the generality  of our result.   Using the  results of
 Elmegreen (1989) (his S function and $s$ are related simply as $s=\rm
 S^{3/5}$), we obtain

\begin{equation}
s\sim 1.1\left(\frac{\rm P_{e}}{10^4\rm \ cm^{-3}\ K}\right)^{13/20}
\left(\frac{\rm M}{10^5\rm \ M_{\odot }}\right)^{-3/10} \left(\frac{\rm T}{100
\rm K}\right)^{3/10} \left(\frac{\mu \rm z}{\rm G_{\circ}}\right)^{3/5}.
\end{equation}

\noindent
It is  clear that unless  M is a  strong function of $\rm  R_{gal}$, a
radially  declining $\rm  P_{e}$ is  the only  other  important factor
besides  $\mu  \rm  z/G_{\circ}$  in determining  $\rm  R_{tr}$.   The
increase of  s with decreasing M  may seem counterintuitive  but it is
due to the  fact that small clouds have  higher average densities than
big  ones  with the  same  boundary  pressure.   This stems  from  the
widespread virialization  of most of the clouds,  which in combination
with  the observed velocity  dispersion vs  cloud size  relation (e.g.
Larson 1981),  yields $\rm n \propto R_{cl}  ^{-\alpha}$ ($\alpha \sim
1$).  In  the metal-rich  part of  a typical spiral,  where CO  can be
relied upon as an H$_2$ gas mass tracer, the bulk of the H$_2$ mass is
distributed in  a cloud  hierarchy with $\rm  M\sim 10^5\  M_{\odot }$
being  typically the  upper limit.   The value  of the  latter  is not
observed to  change with  galactocentric distance (e.g.   Elmegreen \&
Elmegreen  1987) except in  galactic centers  where tidal  effects may
truncate such a cloud hierarchy  at higher minimum densities and cloud
masses (e.g. Maloney \& Black 1988).

 The  measurement of  $\rm P_{e}$,  especially as  a function  of $\rm
R_{gal}$, is not  easy and deducing it by  assuming equipartition with
the random cloud motions (e.g.  Honma  et al.  1995) relies on a known
total gas surface density, which is what needs to be determined.  This
task is easier  for the Galaxy whose disk we  consider typical in that
respect.   In this  case a  pressure probe  sampling its  outer parts,
namely diffuse  CO-luminous molecular clouds,  suggests pressures $\rm
P_{e}(12  kpc)\sim   (1-2)\times  10^4  \rm  \   cm^{-3}\  K$  (Heyer,
Carpenter, \& Snell 2001), which  are higher than expected and similar
to  the  Solar neighborhood  value.   Further  out  pressures of  $\rm
P_{e}(17  kpc)\sim 5\times  10^3  cm^{-3}\ K$  are  typical, with  few
clouds  reaching  up  to  $\sim  10^5\rm  \  cm^{-3}\  K$  (Brand,  \&
Wouterloot  1995).  It  is  straightforward to  show  that since  $\rm
0.25\la z/G_{\circ}\la 1$ (at $\rm  R_{gal}\la 17$ kpc), for $\mu =5$,
such  $\rm P_{e}$  values yield  $s\ga 1$.   Thus the  H$_2$ formation
conditions are fulfilled out to $\sim 2.5\rm R_{e}(\sim 17\rm \ kpc)$.
This  radius  is well  past  the location  of  the  $ ^{12}$CO  J=1--0
brightness peak at $\sim 5$ kpc or any significant $ ^{12}$CO emission
(e.g.   Blitz 1997)  and well  into  the HI  disk.  Interestingly  HII
regions,  another potential pressure  probe, may  also hint  at larger
than expected pressures in the outer Galaxy (Rudolph et al.  1996) and
dwarf irregulars (Elmegreen, \& Hunter  2000).  In both of these cases
low metallicities make $ ^{12}$CO  a poor indicator of H$_2$ gas which
may thus be responsible for the high pressures.

\section{Molecular gas at large galactocentric radii: CO-deficient and warm}

The  arguments  for  favorable  conditions for  an  HI-to-H$_2$  phase
transition presented here  do not and could not  yield a mass estimate
for the associated H$_2$ gas.   Nevertheless the simple fact that this
gas resides untraceable by CO in  the metal-poor parts of the disk can
offer   significant  insights   regarding  its   state.    An  obvious
consequence of  the low metallicities is reduced  dust shielding which
allows even the weaker FUV radiation field at large $\rm R_{gal}$'s to
dissociate the mainly dust-shielded CO  and confine it to much smaller
volumes  while leaving  the  largely self-shielding  H$_2$ intact.   A
typical range of $\rm G_{\circ}\sim 0.3-0.6$ and an $\mu\sim 5$ yields
a minimum H$_2$ formation density  of $\rm n_{\rm f}\sim 5-7\ cm^{-3}$
for  cloud sizes  of $\sim  10-50 $~pc.   Observations  of CO-luminous
molecular clouds yield average densities of $\rm \sim 50-500\ cm^{-3}$
irrespective  of  galactocentric  distance,  hence $\rm  n\sim  5-500\
cm^{-3}$   and   $\rm  z\sim   0.15$   are   appropriate  inputs   for
Photodissociation Region  models (PDR) of  the molecular gas  at large
$\rm R_{gal}$.  Similar modeling  of metal-poor gas has been performed
before but at much higher UV  fields and gas densities than here (e.g.
Pak et  al. 1998).

We investigate the  gas abundance of warm H$_2$ using  a PDR code that
models the chemistry of a diffuse cloud represented as a semi-infinite
slab,  with a  pseudo-time-dependent  multi-point chemical  simulation
(Viti et al.  2001).  The UMIST chemical database (Millar et al. 1997)
was used with  some changes (see Viti et al.   2001). The heating, the
cooling and  CO emission is  as described by  Taylor et al  (1993); we
have however  made the  following changes: i)  the thermal  balance is
achieved  using   Ridder's  method,  a   fast  non-linear  convergence
algorithm,  ii)  the heating  mechanisms  have  been  revised and  now
include also  large grains and  PAHs (Weingartner \& Draine  2001), as
well as UV pumping (not important for low G$_{\circ}$'s and densities)
and chemical heating, iii) the  collisional rates have been updated to
include all  the dominant collisional  partners for such an  ISM phase
(H, H$_2$, He).  All our runs were performed assuming a microturbulent
line formation  with a  typical width  of $\sim $3  km s$  ^{-1}$, and
following Tielens \& Hollenbach 1985 the level populations are updated
from the outside in.  The heating caused by the decay of the turbulent
gas is negligible.





The results are summarized in  Figure 1 where it becomes apparent that
   in diffuse gas of  sub-solar metallicity the CO-deficient H$_2$ gas
   dominates  the cloud.   This  radiation-driven spatial  segregation
   also  maintains a temperature  difference in  which the  pure H$_2$
   phase stays  always warmer than  the CO-luminous one.   This simply
   reflects the different dominant coolants present, namely $ ^{12}$CO
   and its isotopes for the  CO-luminous phase, and CII, H$_2$ for the
   CO-deficient one.  Average temperatures  of $\rm T_{k}\sim 15-20$ K
   are observed for most quiescent CO-luminous molecular clouds in the
   disk (with $\rm T_{k}\sim  30-60$~K found only in star-forming warm
   ``spots'') and  remain almost constant with $\rm  R_{gal}$.  In the
   Galaxy this is beautifully demonstrated by {\it COBE}\ observations
   of CO  J+1$\rightarrow $J (J=0--7)  lines that show  invariant {\it
   relative} strengths suggesting $\rm T_{k}\sim 20$ K over the entire
   disk except the Galactic  center (Fixsen, Bennett, \& Mather 1999).
   This  temperature  is  similar   to  those  of  several  individual
   molecular clouds, some  as far as $\rm R_{gal}\sim  30$ kpc (Digel,
   de Geus,  \& Thaddeus 1994), and  also similar to  that measured by
   {\it COBE}\ for the bulk of the dust in the Galaxy (Sodroski et al.
   1994; Lagache et al.  1998)  and by sub-mm imaging in other spirals
   (e.g.   Alton et  al.   1998).  Such  gas  temperatures are  indeed
   expected for the dust-shielded ($\rm A_{\rm v}\ga~4$) CO-rich H$_2$
   clouds, where CO cooling (Goldsmith, \& Langer 1978; Neufeld, Lepp,
   \& Melnick 1995) lowers $\rm T_{k}$ to that of the concomitant dust
   whose temperature is weakly dependent  to the ambient ISRF and thus
   to~$\rm R_{gal}$.

  For the  unshielded regions  of the molecular  ISM, now  expected to
  contain the  bulk of the H$_2$  gas mass in the  metal-poor parts of
  spiral disks, lack of the  CO coolant will keep $\rm T_{k}>T_{dust}$
  (e.g.   Tielens,   \&  Hollenbach  1985). Its temperature range can 
  be found from

\begin{equation}
\Lambda _{\rm C II}+\Lambda _{\rm H_2}=\Gamma _{\rm ph}+\Gamma _{\rm CR},
\end{equation}

\noindent
where  the cooling  $(\Lambda )$  due to  the CII  and H$_2$  lines is
balanced by  the photoelectric heating from dust  grains $\Gamma _{\rm
ph}$ (e.g.  Tielens \& Hollenbach 1985) and cosmic rays.  Unlike solar
metallicity environments,  the cooling due  to H$_2$ lines may  now be
important and is included.

The  CII emission  emanates  from a  2-level  system, hence  following
Hollenbach,  Takahashi,  \&  Tielens  (1991) and  for  optically  thin
emission it is

\begin{equation}
\Lambda    _{\rm    CII}=1.82\times     10^{-23}    \    \rm    z    n
\left[1+\frac{1}{2}\rm e^{92/\rm T} \left(1+\frac{\rm n _{\rm cr}}{\rm
n}\right)\right]^{-1}\ \rm ergs\ \ s^{-1},
\end{equation}

\noindent
where $\rm n_{\rm cr}\sim 3\times 10^3\ cm^{-3}$ is the transition's
critical density, and all carbon is assumed in the form of CII, mixed
with a fully molecular gas phase at an abundance of $\rm
[C/H]=3\times 10^{-4}\ z$.

For  the expected  range of  densities and  $\rm T_{k}\la  200$  K, (a
typical  upper limit  for  the  Cold Neutral  Medium  (CNM) in  spiral
disks), only the S(0) and S(1)  lines rotational lines of H$_2$ can be
appreciably  excited and  thus contribute  to the  cooling.  Moreover,
since  no  radiative  or  collisional  processes  ``mix''  ortho-H$_2$
(J=odd) and para-H$_2$  (J=even) states we can consider  S(0) and S(1)
practically arising from two different molecules, each approximated by
a 2-level system, therefore

\begin{equation}
\Lambda _{\rm H_2}=2.06\times 10^{-24}\frac{\rm n}{1+\rm r_{op}}\left[
1+\frac{1}{5}\rm e^{\rm 510/T}\left(1+\frac{\rm n _{\rm cr}(0)}{\rm n}\right)\right]^{-1}
\left(1+\rm R_{10}\right) \rm ergs\ s^{-1},
\end{equation}

\noindent
where  $\rm  r_{op}=(n_3+n_1)/(n_2+n_0)$  is the  ortho-to-para  H$_2$
ratio, $\rm n_{\rm cr}(0)\sim 54\rm \ cm^{-3}$ is the critical density
of  the S(0)  transition  and $\rm  R_{10}=\Lambda _{\rm  S(1)}/\Lambda
_{\rm S(0)}$, namely

\begin{equation}
\rm R_{10} = 26.8 r_{op}\left[ \frac{1+\frac{1}{5}\rm e^{510/T}\left(1+\frac{\rm n _{\rm cr}(0)}
{\rm n}\right)}{1+\frac{3}{7}\rm e^{845/T}\left(1+\frac{\rm n _{\rm cr}(1)}
{\rm n}\right)}\right].
\end{equation}

\noindent
The critical density of the S(1) transition is $\rm n _{\rm cr}(1)\sim
10^3\ cm^{-3}$, and this term dominates for the warmest and densest
gas in the chosen parameter space.

We  adopt the  latest photoelectric  heating model  of  Weingartner \&
Draine  (2001) with $\Gamma  _{\rm ph}$  proportionally scaled  by the
normalized  metallicity  factor  z  and  parameters  corresponding  to
carbonaceous  grains (the  last two  models  in their  Table 2).   The
assumed electron abundance  is the average for the  CNM, under similar
set of $\rm (z,  G_{\circ})$, i.e.  $\rm n_{e}/2n\sim 5\times 10^{-4}$
(Wolfire  et al.   1995).  Finally  we include  a heating  factor from
cosmic rays $\Gamma _{\rm  CR}=6.4\times 10^{-28}\rm n$ ergs s$ ^{-1}$
(Goldsmith \&  Langer 1978), which  is $\la 10\%$ of  the photelectric
term.

For $\rm G_{\circ}=0.3-0.6$, $\rm z=0.15$ and $\rm r_{op}=1-3$, both a
diffuse  ($\sim 5-20$  cm$ ^{-3}$)  and  a dense  ($\sim 100-500$  cm$
^{-3}$)  H$_2$ gas  phase satisfy  (4)  with the  former being  warmer
($\sim 50-100$~K) than the latter  ($\sim 20-30$ K).  At least for the
Galaxy, these  solutions can be  further constrained by the  values of
the total pressure  $\rm P_{e}\la (5-20)\times 10^3$ cm$  ^{-3}$ K, at
$\rm R_{gal}\ga 8$ kpc, where

\begin{equation}
\rm P_{e}=n k_{B} T + 56.8 n \sigma_z ^2,
\end{equation}

\noindent
(e.g.  van Dishoeck  \& Black  1986).  The  second term  expresses the
non-thermal part of the pressure due to the macroscopic motions of the
gas, which are assumed of  turbulent origin with $\sigma _z$ being the
one-dimensional  FWHM  of   a  gaussian  distribution.   The  velocity
dispersions  of CO-bright molecular  clouds or  HI in  face-on spirals
yield  $\sigma _z\sim  5$ km  s$  ^{-1}$ (Combes  1999 and  references
therein)  hence  the  aforementioned  constraints on  the  total  disk
pressure correspond to $\rm n\sim 5-15$ cm$ ^{-3}$ (for $\rm T\la 100$
K).  {\it Thus  a metal-poor, warm and diffuse H$_2$  gas in the outer
parts of spiral disks is likely  to exist,} and may indeed be the type
of gas  detected through its S(0)  line with {\it ISO}  in the edge-on
spiral  NGC~891 (Valentijn  \& van  der  Werf 1999),  with a  reported
temperature of $\rm  T\sim 90$ K and a mass that  outweighs that of HI
by factors of $\sim 5-15$.


Several  issues regarding  such a  gas phase  remain open,  namely its
thermal stability, low star formation rates, its exact relation to the
HI gas  distribution and the  dynamical evolution of  such potentially
massive gaseous disks.  Moreover  an enhanced H$_2$ formation rate, if
indeed at  work, may  alter relevant aspects  of the  ``standard'' ISM
picture  (e.g. gas  heating)  that must  be  taken into  account in  a
detailed analysis.   These and other issues  will be the  subject of a
future paper.

\section{Conclusions}

In the  light of mounting evidence  that the standard  method of using
rotational line  emission of  the $ ^{12}$CO  molecule to  trace H$_2$
underestimates its mass at low  metallicities we examined the state of
such  gas in  the disks  of  spiral galaxies  at large  galactocentric
distances.  Our results can be summarized as follows,

\noindent
A high H$_2$ formation rate along with the presence of dust throughout
a typical  HI disk, both  suggested by recent observations,  raise the
possibility of an extended H$_2$ gas phase, well past the one inferred
by the $ ^{12}$CO brightness distribution.

\noindent
This phase is likely to be  warm and diffuse and it may be responsible
for  the  observed  high   total  pressures  at  large  galactocentric
distances  in the  Galaxy.  In  the  case of  NGC 891  such a  massive
gaseous reservoir may have already been detected through its S(0) line
emission.

\acknowledgments

We thank the  anonymous referee for useful comments  that improved the
present work. P. P. P. would  also like to thank Paula Gutierrez Ruiz
for many inspiring conversations that helped this work through.


\clearpage

\begin{figure}
\figcaption{The relative  abundance profiles of H$_2$, H,  CII, CI and
CO for  a one-side  illuminated cloud with  total H nuclei  density of
$\rm n=20\ cm^{-3}$ (see text).   The values of metallicity, FUV field
and formation rate  used are denoted in the upper  or bottom left. The
effect   of  an   enhanced  formation   rate  in   making   the  cloud
H$_2$-dominated within much smaller spatial scales is apparent.}
\plotone{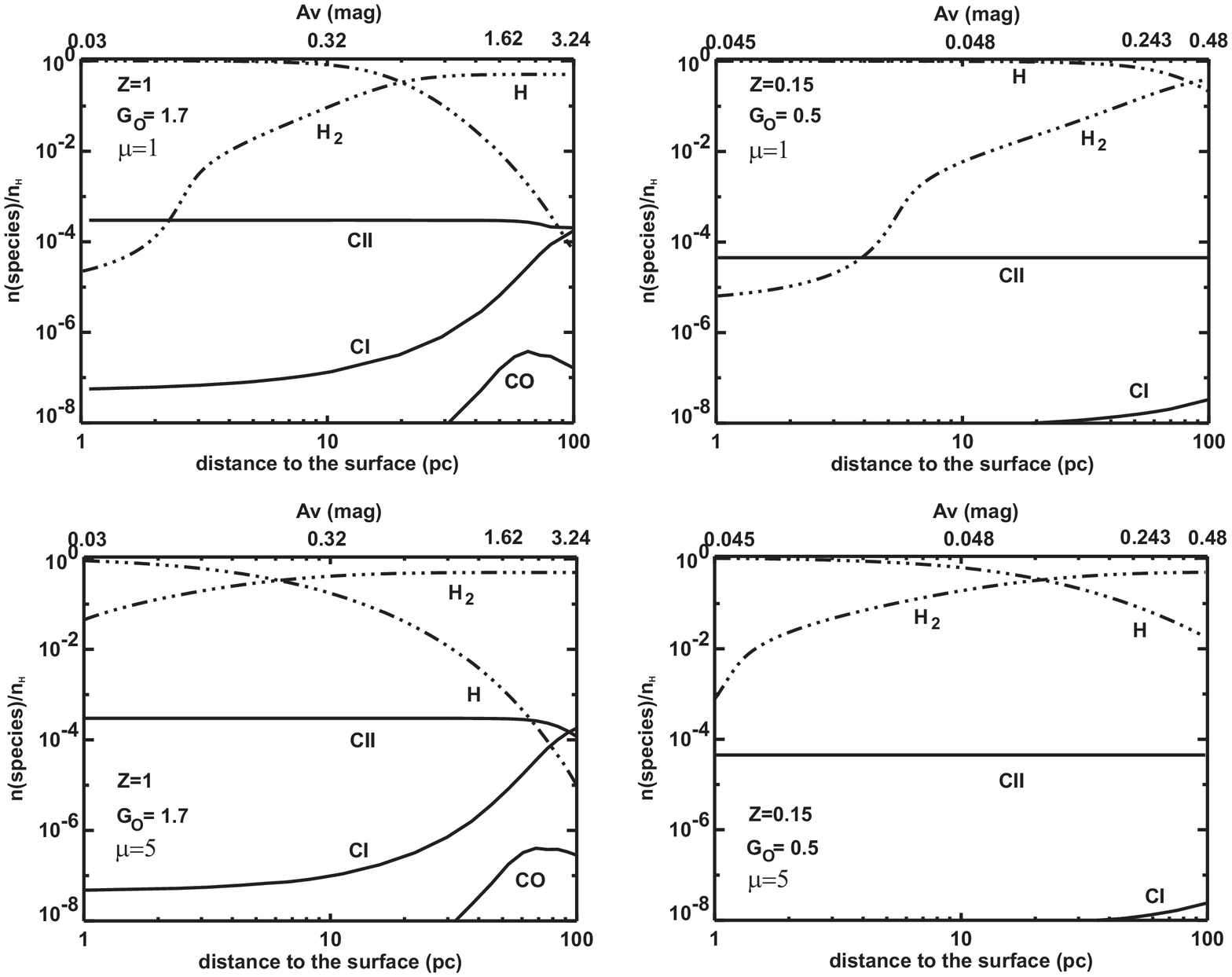}
\end{figure}

\end{document}